\documentclass{iaus}
\usepackage{graphicx,natbib}

\newcommand{\be}{\begin{eqnarray}}
\newcommand{\ee}{\end{eqnarray}}

\newcommand{\Msol}{M_\odot}
\newcommand{\mpl}{m_{\rm p}}
\newcommand{\MJ}{M_{\rm J}}

\newcommand{\mbin}{m_{\rm bin}}
\newcommand{\abin}{a_{\rm bin}}
\newcommand{\ebin}{e_{\rm bin}}
\newcommand{\ikoz}{I_{\rm Koz}}

\newcommand{\ibin}{I_{\rm bin}}

\newcommand{\tkoz}{\tau_{\rm Koz}}

\title[Formation of Close-in Planets in Binaries] 
{Induced Kozai Migration and Formation of  \\ Close-in Planets in Binaries}

\author[Takeda, Kita \& Rasio]   
{Genya Takeda, Ryosuke Kita,
 \and Frederic A. Rasio}

\affiliation{ Dept. of Physics \& Astronomy, Northwestern University, \\ 
2145 Sheridan Road, Evanston, IL USA \\email: {\tt genya@u.northwestern.edu}}

\pubyear{2008}
\volume{253}  
\pagerange{1--7}
\jname{Transiting Planets}
\editors{A.C. Editor, B.D. Editor \& C.E. Editor, eds.}
\begin{document}

\maketitle

\begin{abstract}
Many recent observational studies have concluded that planetary systems commonly exist 
in multiple-star systems.  At least $\sim$20\,\%, and presumably a larger fraction of the known 
extrasolar planetary systems are associated with one or more stellar companions. These stellar 
companions  normally exist at large distances from the planetary systems 
(typical projected binary separations are on the orders $10^2\,$--\,$10^4$AU) and are often faint 
(ranging from F to T spectral types). Yet, secular cyclic angular momentum exchange with 
these distant stellar companions can significantly alter the orbital configuration of the 
planets around the primaries.  One of the most interesting and fairly 
common outcomes seen in numerical simulations is the opening of a large mutual inclination angle 
between the planetary orbits, forced by  differential nodal precessions caused by 
the binary companion. The growth of the mutual inclination angle between planetary orbits 
induces additional large-amplitude eccentricity oscillations of the inner planet due to 
the quadrupole gravitational perturbation by the outer planet. This eccentricity oscillation 
may eventually result in the orbital decay of the inner planet through tidal friction, as previously 
proposed as Kozai migration or Kozai cycles with tidal friction (KCTF). This orbital decay mechanism 
induced by the binary perturbation and subsequent tidal dissipation 
may stand as an alternative formation channel for close-in extrasolar planets. 

\end{abstract}

\firstsection 
\section{Introduction}
As of June 2008, at least $\sim$20\,\% of the $\sim$240 stars hosting planets
are components of multiple-star systems \citep{desidera07,raghavan06}.  
Most of the binaries that harbor planets are very wide, with typical sky-projected 
separations ranging from $10^2$ to $10^4$\,AU.  
Majority of the stellar companions to planetary systems are FGK-type stars, 
but there are also later-type companions, 
including 8 M-dwarfs and one T-type brown dwarf.  These distant, 
and often faint, companions, however, may still play a significant role in  altering
planetary orbits around the primaries on a very long timescale.  As first pointed out by \citet{zucker02},
the two populations of planets, those around single stars and those around components of 
binary systems, seem to exhibit different orbital properties.  
The effect of secular binary perturbations is particularly discernible in the 
distribution of planets' orbital eccentricities that are possibly excited 
by the Kozai mechanism (\S~\ref{kozai}).
For example, a large fraction of planets with highly eccentric orbits ($e > 0.7$) 
are associated with stellar companions.  Also, the population of planets in binaries
shows a higher median eccentricity (0.25) compared to that of the planets around single stars (0.20).
\footnote{observational biases seem to influence the observed eccentricity distribution only marginally; 
see for example \citet{shen08}.}
Another curious trend that was first noticed by \citet{zucker02} and has remained robust still to date
is the population of massive hot Jupiters ($\mpl \sin{i} > 2 \MJ$ and $P < 100\,$days) found exclusively
among binary systems (there are a few recent exceptions such as XO-3\,b or HD\,17156\,b, 
for which the presence of stellar companions has not yet been  ruled out).

Interestingly, secular eccentricity-pumping mechanisms such as the Kozai mechanism
are closely related to the formation of close-in (possibly transiting) planets.   
In addition to the traditional planetary migration theory incorporating planet-disk
interactions, there is a growing attention to an alternative mechanism involving 
circularization of initially eccentric orbits through tidal dissipation in the 
central stars.  There are two major ways a planet can gain a sufficiently large orbital
eccentricity that leads to the tidal circularization and orbital decay: mutual, impulsive
gravitational scattering between planets upon dissipation of the nascent gas
disk \citep{nagasawa08}, and adiabatic secular eccentricity growth induced by the Kozai mechanism
\citep{wu03,fabrycky07}.  The latter scenario, often called Kozai migration or Kozai cycles
with tidal friction (KCTF), has been adopted for single-planet systems in binaries and 
produced models that are consistent with the observations, given certain assumptions 
in the initial conditions. In this study, we investigate the dynamical outcomes of 
multiple-planet systems secularly perturbed by stellar companions and find that Kozai migration
process is significantly enhanced with the presence of additional planetary companions.

\section{Secular Three-body Interactions}
We focus on the secular evolution of hierarchical four-body systems that consist of 
two planets around a component of a wide binary system.  It would help to break up 
the problem into two secular three-body problems in order to provide analytical understandings
of secular four-body interaction: (1) the secular evolution of two planets around the central star 
(nearly coplanar, not necessarily hierarchical), and the secular evolution of 
each planet due to the quadrupole perturbation from the binary (not necessarily coplanar, but highly hierarchical).  
The former has been formulated through
the Laplace-Lagrange Secular Theory, and the latter through the Kozai mechanism.

\subsection{Secular Planet-Binary Interaction}  \label{kozai}
Binaries that contain planets are typically separated by more than 100\,AU.  
In such wide binaries it is likely that the initial orbital angular momentum
of the planet (i.e., the spin-axis orientation of the primary star) is not correlated 
to the orbital angular momentum of the binary \citep{hale94}.
Thus, distribution of the initial inclination of the planetary orbit with respect to the binary plane
is expected to be isotropic. The Kozai mechanism takes its effect only when the initial relative inclination is greater
than the critical Kozai angle $\ikoz = 39.2^\circ$.  In such cases, planet's eccentricity ($e$), 
inclination ($I$), the argument of pericenter ($\omega$), and the longitude of ascending node
($\Omega$), starts to precess secularly all on similar timescales \citep[called Kozai cycles; ][]{kozai62}.
The orbital eccentricity of a planet oscillates, even if it is initially close to zero, 
with a nearly constant amplitude (see for example the red-dashed curve in Figure~\ref{f1.fig}), that is 
to the lowest order a function only of the initial inclination $I_0$ \citep{holman97}:
\be
e_{\rm max} \approx \sqrt{1 - 5/3 \cos^2{I_0}}.
\ee
The timescale of the eccentricity oscillation (and the evolution of other orbital elements)
is determined as \citep{kiseleva98}
\be
\tkoz \approx \frac{2}{3 \pi}\frac{m_0 + \mpl + \mbin}{\mbin} \left(1 - \ebin^2 \right)^{3/2}, \label{tkoz}
\ee
where the subscripts 0 and $bin$ denote the primary and the companion stars, respectively.

During Kozai cycles, a planetary orbit is slowly and repeatedly torqued by the binary,
keeping the orbital energy and thus the semi-major axis constant.
A cumulative effect of the torque results in the secular apsidal precession of the planet,
i.e., $\omega_{\rm p}$ precesses on the timescale $\tkoz$.  In a typical planetary system 
there are other sources of perturbations that also induce secular apsidal precession 
of the planet, such as the general-relativistic effect caused by the central star, or 
secular perturbations from other planetary companions.  If any of these extra perturbations 
induces apsidal precessions on the timescale smaller than $\tkoz$, the torque from the binary companion secularly
averages out to zero, and the planetary orbit will remain nearly circular.

\subsection{Secular Mutual Interaction between Planets}
Analytical solutions for the long-term evolution of a pair of planets around a star 
due to their mutual gravitational interaction is formulated through the second-order 
Laplace-Lagrange secular theory \citep[L-L theory hereafter;][]{brouwer61,murray99}.  
The L-L theory also yields secular evolutionary timescales of the planets.
The two eigenfrequencies  $g_+$ and $g_-$ derived from the L-L theory correspond to 
the apsidal precession timescales, $\tau_{\omega,\pm} \approx 2 \pi / g_{\pm}$. 
\footnote{In certain limits the apsidal evolution is described as a nonlinear combination 
of multiple eigenmodes, and this simple relation between the eigenfrequency and the apsidal
precession frequency does not hold.  See \citep{takeda08}}  This apsidal evolution 
timescale derived from the L-L theory can be compared to the Kozai timescale $\tkoz$ (Eq.~[\ref{tkoz}])
to determine whether the Kozai cycles   are suppressed or not.  
If, for instance,  a double-planet system in a binary has initial configuration such that  
$\tau_{\omega, 2} < \tau_{\rm Koz, 2}$, then the Kozai cycles on the outer planet 
will be entirely suppressed, and the binary companion will play no significant role in 
the secular evolution of the outer planet.

The L-L theory also yields {\em one} eigenfrequency $f$ which scales with the coupling
strength of the planetary ascending nodes \footnote{If an additional quadrupolar potential
such as stellar oblateness is introduced, there will be two eigenfrequencies in the 
$I$-$\Omega$ space.  See \citet{murray99}.}.  This also translates to the nodal 
precession timescale of a double-planet system as $\tau_\Omega \approx 2 \pi / f$.

\section{Non-rigid Evolution of Planetary Orbits in Binaries}

Newly formed planets emerging from the circumstellar gas disk are expected to orbit
in a common plane (along the stellar equator).  However, 
in the quadrupolar gravitational potential of the binary, this coplanarity can be
broken as the planets evolve \citep{takeda08}.
If the nodal coupling of the planets (scales with $\tau_\Omega$) is strong compared to the nodal perturbation caused
by the binary companion (scales with $2\tkoz$), planetary orbits can evolve rigidly, maintaining near-coplanarity.  
On the other hand, if the nodal coupling between the planets is weak compared to the nodal perturbation
introduced by the stellar companion, a large mutual inclination angle starts to grow between the planetary orbits,
thereby inducing Kozai cycles to the orbit of the inner planet.

\begin{figure}[h]
\begin{center}
 \includegraphics[width=0.6\hsize]{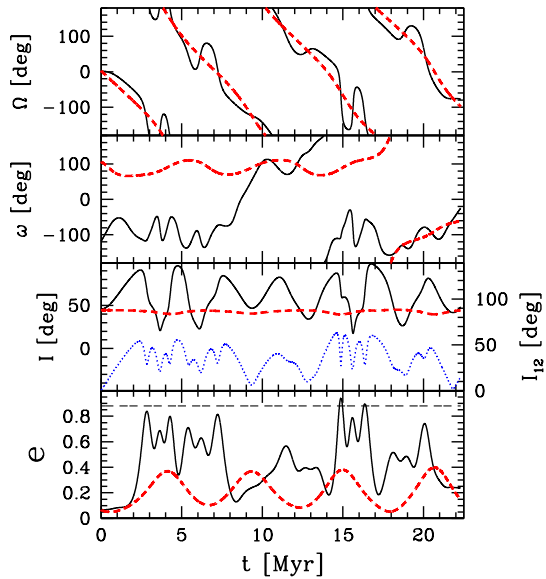} 
    \caption{Numerically integrated evolution of a non-rigid double-planet system in a binary.  
	The system evolved from the following initial conditions: for the inner planet (solid black curve),  
	$m_1 = 0.02 \MJ$, $a_1 = 0.9\,$AU and $I_1 = 45^\circ$; for the outer planet 
	(dashed red curve), $m_2 = 0.6 \MJ$, $a_2 = 12.7\,$AU and $I_2 = 45^\circ$; 
	and for the binary companion,  $\mbin = 0.7 \Msol$, $\abin = 803$\,AU and $\ebin = 0.78$.  
	The binary  orbit remains at $\ibin = 0^\circ$.  The dotted blue curve represents 
	the relative inclination angle $I_{12}$  between the two planets (scale shown on the right axis).  
	At around $t=14.5\,$Myr, the inner planet's pericenter enters inside 0.1\,AU of the central star 
	(horizontal dashed line in the bottom panel) and remains there for $\sim2\times10^5$yr.  }

   \label{f1.fig}
\end{center}
\end{figure}

In Figure~\ref{f1.fig}, an initially zero mutual inclination ($I_{12}$) 
between the planetary orbits grows periodically
to as large as $\sim 60^\circ$.  As soon as the mutual inclination between the planets reaches
$40^\circ$ at $t\sim2\,$Myr, the inner planet's orbital eccentricity $e_1$ starts to oscillate
chaotically, reaching as high as $\sim$0.95.  This growth of $I_{12}$ 
is solely due to the differential nodal precession 
of the planetary orbits introduced by the stellar companion \citep{takeda08}.  
It is almost independent of how inclined initially the planetary orbits are 
with respect to the binary plane, unlike the three-body Kozai mechanism.
In four-body systems, the orbital eccentricity of the inner planet can still grow 
to a very large value even if the planet's initial inclination is below 
the critical Kozai angle.

In an opposite case in which the nodal coupling of the planetary orbits is sufficiently
strong compared to the nodal precession of the outer planet induced by the stellar companion,
planetary orbits secularly rotate together, remaining on a common plane.  In such dynamically-rigid
configuration, $I_{12}$ and $e_1$ remain small throughout the evolution.

\section{Kozai Migration and Formation of Close-in Planets in Binaries}
In dynamically non-rigid systems inner planet's eccentricity secularly oscillates
due to the perturbation from the outer planet.  During these Kozai cycles,
without any dissipative mechanism the semi-major axis $a_1$ of 
the inner planet remains constant throughout the evolution.  This means that 
 at each eccentricity maximum, the planet's pericenter distance $q_1 = a_1 (1 - e_1)$ 
reduces greatly.  It is known that in the regime where $q_1 \lesssim 0.1\,$AU, 
strong tidal interaction with the central star leads to damping of 
the orbital eccentricity and the semi-major axis, and the planetary orbit migrates
toward the central star until it eventually halts at a tight, nearly circular orbit 
\citep{kiseleva98,wu03,fabrycky07}.  For example, in the example in Figure~\ref{f1.fig}
the inner planet is expected to start decaying its orbit around $t\approx15\,$Myr if
tidal dissipation is included in the calculation.

This so-called Kozai migration scenario has been adopted to produce some dynamical models
that are consistent with  observations \citep[e.g., HD\,80606\,b; ][]{wu03}, given 
relatively fine-tuned initial conditions.  The difficulty is to achieve such a  
small pericenter distance during the Kozai cycles that the tidal dissipation significantly
drains the orbital energy from the planet. 
Thus these dynamical models require very large initial relative inclination angles,
almost  $90^\circ$.  Such a large  angle is not unrealistic, 
but not likely to be applicable to many systems.

\begin{figure}[ht]
\begin{center}
 \includegraphics[width=0.7\hsize]{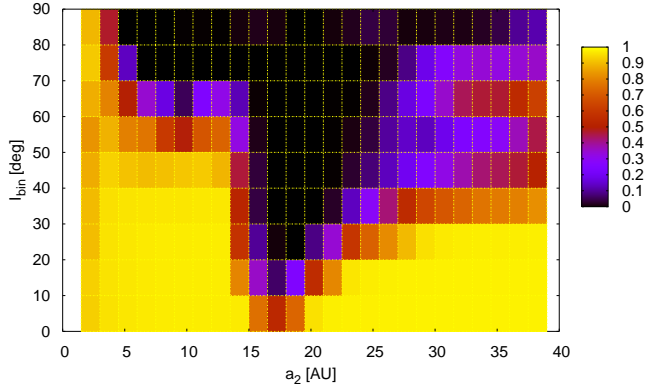} 
    \caption{The minimum pericenter distance in AU reached by the inner planet within 500\,Myr
	of numerical integrations with different sets of $\ibin$ and $a_2$.  
	Other orbital parameters and masses are fixed as $m_0 = \mbin = 1.0\Msol$, $m_1 = 0.1 \MJ$, $m_2=1.0\MJ$, $a_1=1\,$AU,
	 $\abin = 750\,$AU, and $\ebin =0.5$.  The darker-shaded area corresponds to the smaller 
	 pericenter distance achieved by the inner planet.  Planets that reach pericenter approach
	 smaller than $\sim$0.1\,AU are subject to tidal dissipation and expected to start migrating toward
	 the central star.}
   \label{f2.fig}
\end{center}
\end{figure}

However, as explained in the previous section, in a more realistic planetary system containing
more than one planet, such a tight requirement on the initial inclination angle is no longer
necessary.
Figure~\ref{f2.fig} shows the distribution of minimum pericenter distances achieved 
by the inner planet during 500\,Myr of numerical integration with different sets of initial 
$a_2$ and $\ibin$.  The parameter space is divided into two evolutionary classes depending on the location
of the outer planet; in the region where $a_2 \lesssim 15\,$AU, 
the planetary system is dynamically rigid, and the inner planet's eccentricity remains
small (pericenter remains large).  The only exceptions in this region are when the planetary orbits are inclined
from the binary by more than $\sim60^\circ$, in which case close encounter
or possible orbital crossing between the planets induced by the large-amplitude Kozai cycles
of the outer planet impulsively excites $e_1$ to large values.  
If, on the other hand, $a_2 \gtrsim 15\,$AU initially, then the system is dynamically 
non-rigid, and even if the planetary orbits and the binary plane are initially nearly 
coplanar, additional Kozai cycles will significantly excite the orbital eccentricity of the 
inner planet.

Such a numerical coverage of $a_2\,$--\,$\ibin$ space shows similar morphology even if different
initial binary parameters and planetary masses are selected; there is always a 
critical orbital separation $a_{2, \rm crit}$ that separates dynamically-rigid and 
non-rigid evolutions.  As illustrated in Figure~\ref{f2.fig} the formation of 
close-in planets through tidal circularization of eccentric orbits should be particularly 
efficient if the outer planet is initially located beyond $a_{2, \rm crit}$.  This boundary
$a_{2, \rm crit}$ can be analytically determined by simple formulae, 
given the initial orbital parameters of the system \citep{takeda08}.

\section{Properties of Close-in Planets Formed via Enhanced Kozai Migration}
As demonstrated in the previous chapter, a stellar companion can effectively 
stir up planetary systems from a large distance and excite the eccentricity
of the innermost planet by propagating its perturbation inward through planets 
on a reasonable timescale.  Once the planet's eccentricity 
becomes sufficiently large and the pericenter distance sufficiently small, the planet
begins to migrate toward the central star until it settles on a tight, circular orbit.
It is possible to identify close-in planets in binaries that formed via 
such an ``induced Kozai migration'' scenario from their final orbital properties.

The leading hypothesis for the planet migration mechanism involves the loss of angular momentum through
 interaction with the protoplanetary gas disk \citep[e.g., ][]{goldreich03}.  If the 
planet--disk interaction is the common migration channel for extrasolar planets, 
majority of the close-in planets are expected to be found on orbits along the stellar equator.
On the other hand, if a planet migrates via the combined effect of the Kozai cycles and 
tidal dissipation, the angular momentum of the final planetary orbit 
should be largely misaligned with respect to the stellar spin axis since the planet's inclination 
evolves during the Kozai cycles
\footnote{Theoretically, orbital inclination can be also damped toward the stellar equator 
through tidal dissipation.  However, it usually takes longer than the estimated 
age of the planetary system; see \citet{hut81,winn05}.}.
The sky-projected angle $\lambda$ of spin-orbit misalignment is now measurable for transiting planets
through the Rossiter--McLaughlin (RM) effect \citep[e.g., ][]{winn05}.  Among the first 8 extrasolar planets with 
available RM measurements, two of them show large spin-orbit misalignments; HD\,17156b 
\citep[$\lambda = 62^\circ\pm25^\circ$; ][]{narita08} and XO-3b \citep[$\lambda = 70^\circ\pm15^\circ$; ][]{hebrard08}.
Although these results still need to be verified by follow-up observations, once the spin-orbit misalignment
is confirmed they will be the first strong candidate transiting systems that formed through
Kozai migration. 

Another curious observational trend is the presence of massive
close-in planets  found predominantly among binaries.
A possible indication of this trend is that massive planets somehow do not migrate efficiently 
through planet-disk interaction.  If that is true, then Kozai migration triggered by binary perturbations
would stand as a strong alternative to form close-in planets with a wide range of masses.
Another possibly common migration channel, planet--planet scattering during the early stage
of planet formation \citep{chatterjee08,nagasawa08}, also faces a significant challenge
in migrating massive planets into tight orbits because it is usually the lighter planets that
get scattered inward.  The Kozai mechanism combined with tidal
damping so far is the only process that does not depend strongly on planetary masses,
consistent with the exclusive presence of massive ones in binaries.
This is another reason to suspect that the two transiting planets 
HD\,17156b \citep[$\mpl = 3.09^{+0.22}_{-0.17}\MJ$; ][]{gillon08}
and XO-3b \citep[$\mpl = 11\,$--\,13$\MJ$; ][]{johns-krull08,winn08} are the outcomes of 
Kozai migration, as both planets are fairly massive. 

The orbital migration process proposed in this study requires at least four or more bodies 
in a system, including two stars and two (or more) planets.  
Alternatively, for a handful of known close-in single planets in binaries, 
detections of secondary planets will significantly constrain the dynamical origin of those systems.
Note that the induced Kozai migration is more common in hierarchical planetary systems
where the second planet orbits beyond a critical radius $a_{2, \rm crit}$ such that the 
planetary orbits evolve non-rigidly.   Unfortunately, detection of additional planets 
beyond a few AU will take several years or decades of continuous radial-velocity monitoring.
Once the presence of additional planetary companions in those systems are ruled in or out, 
however, the dynamical history of known close-in planets can be significantly constrained.

\vspace{0.5cm}
This work was supported by NSF Grant AST-0507727.


\end{document}